\documentclass[conference]{IEEEtran}
\IEEEoverridecommandlockouts
\usepackage{cite}
\usepackage{amsmath,amssymb,amsfonts}
\usepackage{algorithmic}
\usepackage{graphicx}
\usepackage{textcomp}
\usepackage{xcolor}
\def\BibTeX{{\rm B\kern-.05em{\sc i\kern-.025em b}\kern-.08em
    T\kern-.1667em\lower.7ex\hbox{E}\kern-.125emX}}

\usepackage{color,soul}

\allowdisplaybreaks

\begin{document}

\title{ISO and DSO Coordination: A Parametric Programming Approach
\thanks{This work is supported by Power Systems Engineering Research Center.}
}

\author{\IEEEauthorblockN{Mohammad Mousavi}
\IEEEauthorblockA{\textit{Arizona State University} \\
mmousav1@asu.edu}
\and
\IEEEauthorblockN{Meng Wu}
\IEEEauthorblockA{\textit{Arizona State University} \\
mwu@asu.edu}
}

\maketitle

\begin{abstract}
In this paper, a framework is proposed to coordinate the operation of the independent system operator (ISO) and distribution system operator (DSO) to leverage the wholesale market participation of distributed energy resources (DERs) aggregators while ensuring secure operation of distribution grids. The proposed coordination framework is based on parametric programming. The DSO builds the bid-in cost function based on the distribution system market considering its market player constraints and distribution system physical constraints including the power balance equations and voltage limitation constraints.  The DSO submits the resulting bid-in cost function to the wholesale market operated by the ISO. After the clearance of the wholesale market, the DSO determines the share of its retail market participants (i.e., DER aggregators). Case studies are performed to verify the effectiveness of the proposed method.
\end{abstract}


\section{Introduction}
US Federal Energy Regulatory Commission issued Order No. 2222 to promote wholesale market competition by leveraging the market participation of distributed energy resources (DERs) \cite{ferc_2222}.  Integrating numerous small DERs into today's wholesale market causes challenges for the independent system operators (ISOs) as 1) it imposes complexity and computational burden; and 2) it could cause distribution grid voltage/thermal violations if the aggregator-controlled DERs are not properly monitored by system operators. An effective solution to this problem is considering the distribution system operator (DSO) which runs the retail market to coordinate the DERs market participation while assuring the secure/reliable operation of the distribution grid \cite{mousavi2021dso}. However, there is a need for a coordination framework between the ISO and the DSO.  

Recent works studied the ISO-DSO coordination problem \cite{chen2021distribution,liu2020integrating,moret2020loss,liu2020accurate,renani2017optimal, wang2021real,yin2021stochastic,haider2020toward,bragin2021tso,khodadadi2020non, hassan2018energy}. In \cite{chen2021distribution}, a bi-level optimization model is proposed for DSO market clearing and pricing considering ISO-DSO coordination. The clearing conditions for the DSO and ISO markets are proposed in the upper-level and lower-level problems, respectively. The problem is converted to mixed-integer linear programming via an equilibrium problem with equilibrium
constraints (EPEC) approach. However, the proposed model is hard to solve for large systems. In \cite{liu2020integrating}, a feasible region-based approach is proposed for DERs' wholesale market integration considering the physical constraints of the DSO-operated distribution grid. However, the market settlement procedure between the DSO and ISO is not proposed. In \cite{moret2020loss}, an extension of the decentralized market framework is proposed to consider loss allocation and its impact on the market outcome. However, the decentralized market framework is not compatible with the current market structures. In \cite{liu2020accurate}, a multi-port power exchange model is proposed to integrate numerous DERs into the wholesale market considering the distribution grid physical constraints. However, the market settlement procedure is not proposed. In \cite{renani2017optimal}, a transactive market framework starting from the ISO to the DSO is proposed. The DSO runs the transactive market using an iterative method. However, the convergence of the proposed method is not guaranteed. In \cite{wang2021real}, a Nash bargaining-based method is proposed for the market-clearing process and the ISO-DSO coordination. However, there is no guarantee that the proposed algorithm converges especially when the number of DSOs increases. In \cite{haider2020toward}, a distributed optimization algorithm is proposed for modeling the DSO retail market considering energy and ancillary services. However, the DSO's impact on wholesale market clearing is not considered. In \cite{yin2021stochastic}, a three-stage unit commitment (UC) is proposed for the transmission-distribution coordination. A convex AC branch flow model is proposed to handle the distribution grid physical constraints. In \cite{bragin2021tso}, the optimal operation and coordination of the ISO-DSO is proposed. A decomposition algorithm is proposed and the original problem is decomposed to ISO and DSO sub-problems. In \cite{khodadadi2020non}, a non-cooprative game approach is proposed for ISO-DSO coordination in which they optimize their own operational costs. In \cite{hassan2018energy} a bi-level optimization model is proposed for the energy storage sizing and siting problem in the DSO-ISO coordination framework. The approaches in \cite{yin2021stochastic, bragin2021tso, khodadadi2020non, hassan2018energy} are hard to solve for large systems.    

This paper proposes an ISO-DSO coordination framework based on parametric programming. The DSO builds the bid-in cost function (submitted to the ISO), considering its retail market participants' offering prices and their operational constraints as well as physical constraints of the distribution grid including power balance equations and voltage limitation constraints. To our best knowledge, this is the first attempt which shows the parametric programming based DSO offers optimal interactions with existing ISO markets.
Different from existing approaches facing computational difficulties for large-scale ISO-DSO coordination, this work could lead to a coordinated ISO-DSO market clearing procedure which is computationally efficient and scalable toward large-scale systems with many DSOs and numerous DER aggregators. This work is an extension to our recent work \cite{mousavi2021dso} to present a coordination framework for the DSO and ISO which is practical with the current wholesale market structures.

\section{DER Market Participation Framework}
\subsection{Direct Participation of the DERs in the ISO Market}
This section presents the ideal case for DER market participation. This ideal case assumes the DERs participate in the ISO's wholesale market directly, and the ISO considers not only transmission-level operating constraints but also distribution grid physical constraints to ensure transmission and distribution security operations, since the DERs are located in the distribution grid. It is assumed that the ISOs have revised their tariffs such that DERs can participate in the wholesale market under one of the participation models.
A unified formulation for the security constrained UC and economic dispatch (ED) problem of this ideal case is as follows:
\begin{align}\label{equ.1}
	\text{Min}_{x,q} \,\,\,\,\,\,&\sum_{t \in T} \sum_{i \in N} c_{i,t}(x_{i,t-1}, x_{i,t}, q_{i,t})\\
	\begin{split}\label{equ.2}
	\text{s.t.} \,\,\,\,\,\,\,\,\,\,\,& (x,q)\in S^{Tra}\\
	& (x,q)\in S^{Dis}\\
	& (x_{i,t-1}, x_{i,t}, q_{i,t-1}, q_{i,t}) \in S_i^{player}, \forall i,t\\
	& x_{i,t} \in \{0,1\}, q_{i,t} \in \mathbb{R}^1 , \forall i,t\\
	& c_{i,t}(x_{i,t-1}, x_{i,t}, q_{i,t}): \mathbb{R}^3 \mapsto \mathbb{R}^1, \forall i,t\\
	\end{split}
\end{align}
where $t$ ($T$) and $i$ ($N$) are the index (set) for the operating timespan and the market participants (generators/aggregators), respectively; $x_{i,t}$, $q_{i,t}$, and $c_{i,t}(x_{i,t-1}, x_{i,t}, q_{i,t})$ are the binary UC decision variable (start-up/shut-down), continuous ED decision variable (dispatched power), and bid-in cost function (with UCED decisions) of market participant $i$ at time $t$, respectively; $x$ and $q$ denote the vectors of $x_{i,t}$ and $q_{i,t}$ for $t$ $\in$ $T$ and $i$ $\in$ $N$, respectively; $S^{Tra}$, $S^{Dis}$, and $S^{player}_i$ denote the search space defined by the system-wide transmission grid constraints, system-wide distribution grid constraints, and operating constraints of the market participant $i$, respectively. 

This is the ideal case for DERs' wholesale market participation. However, implementing this procedure is not possible for ISOs since 1) it adds many variables/constraints to the ISO problem from the distribution grids, making the ISO problem computationally expensive; and 2) it increases the ISO's burden on modeling the distribution-level constraints in its market clearing problem, while the distribution-level models/constraints are currently not available to ISOs.      
\subsection{Market Participation of the DERs through DSO and ISO Coordination Framework}
This section presents our proposed ISO-DSO coordination framework. This framework decomposes the ideal case in Section II.A into an ISO sub-problem and several DSO sub-problems (one for each distribution system). This decomposition reduces the ISO's modeling and computation burdern by 1) considering distribution-level operating security in the DSO sub-problems; 2) considering distribution-level variables/constraints and optimization computations in the DSO sub-problems; and 3) introducing minimal changes to the existing ISO market structures. If locational marginal pricing (LMP) is adopted by the ISO and DSO markets, the market clearing outcomes of this ISO-DSO coordination framework are identical to those of the ideal case in Section II.A.

Each DSO is defined as a mediator which gathers offers from the DER aggregators to submit an aggregated bid to the wholesale energy market. The DER aggregators submit their offers to the DSO. The DSO gathers these offers and runs the retail market to build an aggregated offer to participate in the ISO wholesale market. Considering the DSOs as wholesale market participants, the wholesale market (ISO) security constrained UCED problem is as follows: 
\begin{align}\label{equ.3}
	\text{Min}_{x,q} \,\,\,\,\,\,&\sum_{t \in T} \sum_{i \in N_{gen}\cup N_{dso}}c_{i,t}(x_{i,t-1}, x_{i,t}, q_{i,t})\\
	\begin{split}\label{equ.4}
	\!\!\!\!\!\!\!\!\!\!\!\!	\text{s.t.} \,\,\,\,\,& (x,q)\in S^{Tra}\\
		& (x_{i,t-1}, x_{i,t}, q_{i,t-1}, q_{i,t}) \in S_i^{gen}, \forall i \in N_{gen}, \forall t\\
		& (x_{i,t-1}, x_{i,t}, q_{i,t-1}, q_{i,t}) \in S_i^{dso}, \forall i \in N_{dso}, \forall t\\
		& x_{i,t} \in \{0,1\}, q_{i,t} \in \mathbb{R}^1 , \forall i,t\\
		& c_{i,t}(x_{i,t-1}, x_{i,t}, q_{i,t}): \mathbb{R}^3 \mapsto \mathbb{R}^1, \forall i,t\\
	\end{split}
\end{align}
where $N_{gen}$ and $N_{dso}$ are the set of all generaors and DSOs in the wholesale market, respectively; $S_i^{gen}$ and $S_i^{dso}$ denote the search space defined by operating constraints of individual generators and DSOs, respectively; $N = N_{gen} \cup N_{dso}$.

Each DSO submits its bid-in cost function to the ISO's UCED problem in (\ref{equ.3})-(\ref{equ.4}), following the ISO-defined non-convex cost function structure $c_{i,t}^{dso}(x_{i,t-1},x_{i,t},q_{i,t})$. Considering aggregator controlled DERs do not have start-up/no-load costs or binary UC decision variables, the bid-in cost function of aggregator $j$ within DSO $i$ at time $t$ reduces to $c_{i,j,t}^{agg}(q_{i,j,t}^{agg})$, which is convex (piecewise linear/quadratic in many markets), where $q_{i,j,t}^{agg}$ is the bid-in power quantity of this aggregator to the DSO. The bid-in cost function of the DSO $i$, $c^{dso}_{i,t}(q^{dso}_{i,t})$ to be submitted to ISO (where $q^{dso}_{i,t}$ is the bid-in power quantity of this DSO to the ISO), is determined by following optimization problem (for single-period DSO markets):
\begin{align}\label{equ.5}
	\begin{split}
	c_{i,t}^{dso}(q_{i,t}^{dso})=\text{Min}_{q^{agg}} &\sum_{j \in {DSO}_i}c_{i,j,t}^{agg}(q_{i,j,t}^{agg})\\
		\text{s.t.} \,\,\,\,\,\,\,& q_{i,t}^{dso}\le \sum_{j \in DSO_{i}}q_{i,j,t}^{agg}\\
		& q_{i,j,t}^{agg} \in S_{i,j}^{agg}, \forall j \in DSO_i\\
		&  q^{agg} \in S^{Dis}_i
	\end{split}
\end{align}
where $DSO_i$ is the set of all aggregators in $i^{th}$ DSO;  $S^{agg}_{i,j}$ is the search space defined by operational constraints of individual aggregators within each DSO; $S^{Dis}_i$ is the search space defined by the physical constraints of each DSO (i.e., the distribution system); $q^{agg}$ is the vector of $q_{i,j,t}^{agg}$ for $j \in DSO_{i}$. 

For a single-period DSO market, Equation (\ref{equ.5}) is a parametric convex optimization problem parameterized by a single parameter $q_{i,t}^{dso}$, since its objective function is sum of convex bid-in cost functions from aggregators, and its constraints are all linear. The optimal solution of (\ref{equ.5}) is a function of parameter $q_{i,t}^{dso}$ which
is the bid-in cost function of $i^{th}$ DSO, $c_{i,t}^{dso}(q_{i,t}^{dso})$. 
Based on approximate multi-parametric convex programming \cite{bemporad2006algorithm}, the optimal bid-in cost function from DSO to ISO, $c_{i,t}^{dso}(q_{i,t}^{dso})$, is also a convex function of parameter $q_{i,t}^{dso}$. If the aggregator bid-in cost functions are (piecewise) linear (following the cost function structure in existing ISOs), this problem reduces to a parametric linear optimization. Based on theories of multi-parametric linear programming \cite{borrelli2003geometric, gal1972multiparametric}, the resulting $c_{i,t}^{dso}(q_{i,t}^{dso})$ is also (piecewise) linear, following the cost function structure in many existing
ISO markets. The optimal outcomes of (\ref{equ.5}) determines: 1) the optimal bid-in cost function $c_{i,t}^{dso}(q_{i,t}^{dso})$ submitted from DSO to ISO (the minimal operating cost for DSO to offer $q_{i,t}^{dso}$ MW of generation/consumption in the ISO market); and 2) the optimal dispatch of total DSO
generation/consumption $q_{i,t}^{dso}$ among all aggregators $q_{i,j,t}^{agg}$ to achieve minimal operating cost. Retail LMPs can be obtained by the dual model (not discussed in this paper).
If a multi-period DSO market is considered, this problem generalizes to a multi-parametric convex optimization problem and all the above discussions are still valid.

This convex (multi)-parametric-programming-based retail energy dispatch can be solved by existing multi-parametric programming solvers \cite{mpt,Lofberg2004,oberdieck2016pop}. If single-period market clearing is considered (currently implemented by many real-world ISOs, as shown in (\ref{equ.5})), this problem boils down to a convex parametric programming problem parameterized by a single parameter. To solve this single-period DSO market clearing problem, we have adopted sensitivity analysis procedure, in which we gradually adjust $q_{i,t}^{dso}$ by a pre-defined small step size and solve the optimization in (\ref{equ.5}) at each step to obtain $c_{i,t}^{dso}(q_{i,t}^{dso})$. The range for adjusting $q_{i,t}^{dso}$ is determined by upper/lower generation limits of individual DER aggregators.

The parametric programming in (\ref{equ.5}) is further expanded to obtain detailed formulation for the DSO market. The bid-in cost function of each DSO is determined by solving (\ref{equ.6})-(\ref{equ.14}): 
\begin{align}\label{equ.6}
		c^{dso}(P^{dso})= Min & \sum_{g \in G}\sum_{b\in B}P_{g,b}\pi_{g,b}-\sum_{d \in D}\sum_{b\in B}P_{d,b}\pi_{d,b} 
\end{align}
\vspace{-18pt}
\begin{align*}
\hfilneg \text{s.t.} \hspace{9000pt minus 1fil}
\end{align*}
\vspace{-18pt}
\begin{align}
	\begin{split}\label{equ.7}
 & \sum_{d\in D}\sum_{b\in B}H_{n,d}P_{d,b}+H^{sub}_nP^{dso}+L^P_n\\
		&-\sum_{g\in G}\sum_{b\in B}H_{n,g} P_{g,b}+\sum_{j\in J}Pl_{j} A_{j,n} =0;\hspace{3mm} \forall n \in N \\
	\end{split}\\
	\begin{split}\label{equ.8}
&\sum_{d\in D}\sum_{b\in B}H_{n,d}P_{d,b} tan\phi_{d}+H_{n}^{sub}Q^{dso}+L_{n}^{Q}\\
& -\sum_{g\in G}\sum_{b\in B}H_{n,g} P_{g,b}tan\phi_{g}+\sum_{j\in J}Ql_{j} A_{j,n} =0 ;\forall n \in N 
\end{split}\\
	& 0 \le P_{g,b} \le P_{b,g}^{max};\hspace{3mm}\forall b \in B, \forall g \in G\label{equ.9}\\
	& 0 \le P_{d,b} \le P_{d,g}^{max};\hspace{3mm}\forall b \in B, \forall d \in D\label{equ.10}\\
	\begin{split}\label{equ.11}
		& U_{m}=U_{n}-2(r_{j} Pl_{j}+x_{j} Ql_{j} );\hspace{3mm}\forall m\in N,\\
		&\hspace{10mm}\forall n \in N,\, C(m,n)=1 ,\, A(j,n)=1 \\
	\end{split}\\
	& \underline{U} \le U_{n} \le \overline{U} ;\hspace{3mm}\forall n \in N \label{equ.12}\\
	& -Pl^{max} \le Pl_{j} \le Pl^{max};\hspace{3mm} \forall j\in J  \label{equ.13}\\
	& -Ql^{max} \le Ql_{j} \le Ql^{max};\hspace{3mm} \forall j\in J  \label{equ.14}
\end{align}
where $t$ and $T$ are the index and set for the entire operating timespan; $g$ and $G$ are the index and set for all generating aggregators; $d$ and $D$ are the index and set for all demand response aggregators; $b$ and $B$ are the index and set for all production/demand blocks; $j$ and $J$ are the index and set for all lines; $n$ and $N$ are the index and set for all nodes; $P^{dso}$ is the DSO's aggregated offers (in MW) to ISO market; $P_{g,b}$ are $P_{d,b}$ are energy offer submitted by the generating aggregators and demand response aggregators, respectively with corresponding prices $\pi_{g,b}$ and $\pi_{d,b}$, respectively; $H_{n,d}$, $H_{n,g}$, and $H_{n}^{sub}$ are mapping matrix of generating aggregators, demand response aggregators, substation to node $n$, respectively; $Pl_j$ and $Ql_j$ are the active and reactive power of branch $j$, respectively; $A_{j,n}$ is the incidence matrix of branches and nodes;  $\phi_g$ and $\phi_d$ are the phase angle of the generating aggregators and demand response aggregators, respectively; $Q_n^D$ is the reactive power of the firm load at each node; $L^P_n$ and $L^Q_n$ are the active and reactive power load at each node; $P^{max}_{g,b}$ and $P^{max}_{d,b}$ are the maximum production/consumption at each block of the generating aggregators and demand response aggregators, respectively; $U$ is the square of voltage of each node; $\underline{U}$ and $\overline{U}$ are the square of minimum and maximum permitted voltage values, respectively; $r_j$ and $x_j$ are resistance and reactance of the branches; $Pl^{max}$ and $Ql^{max}$ are maximum active and reactive power of branches.

Equation (\ref{equ.6}) defines the objective function of the DSO problem to minimize the total system cost. Equations (\ref{equ.7})-(\ref{equ.8}) define the active and reactive power balance equations, respectively. The produced/consumed power at each block of the generating aggregators and demand response aggregators  are limited by Constraints (\ref{equ.9})-(\ref{equ.10}), receptively. Voltage at each node is defined by (\ref{equ.11}). The minimum and maximum voltage limitations are met by (\ref{equ.12}). Constraints (\ref{equ.13})-(\ref{equ.14}) limit the active and reactive power of each branch, respectively. More details for this DSO problem are in our prior work \cite{mousavi2021dso}.

Note that we need to determine the bid-in cost function for each DSO. Once we provide these cost functions, we can submit them to the ISO. Then ISO will clear the wholesale market and determine the share of each DSO in the ISO market. The merit of this procedure is that the ISO does not need to know the inner (distribution-level) constraints of the DSOs. This means that ISO does not need to consider a lot of variables and constraints to ensure the DERs' wholesale bidding activities do not cause voltage/thermal violations in the distribution grids. If LMP is adopted in the ISO-DSO markets, market clearing outcomes of this framework are the same as those of the ideal case in Section II.A, as the ISO is considering market participation of the small DER aggregators in the wholesale market (through the DSO). This is due to the fact that every share that ISO determines for each DSO lies on the best response function of that DSO (already submitted to the ISO). Hence, the results are the same as those in the ideal case when DERs participate in the ISO market directly. Due to space limitation, mathematical proofs are not included. 

In the DSO problem, a parameter $P^{dso}$ determines the amount of the power imported/exported from/to the ISO. In other words, the DSO is considered as a unit that is going to determine the cost function or demand function based on its generating units and demand response units as well as physical constraints of the distribution network. Indeed, the resulting cost function determines the true value cost of the energy consumed or produced in the distribution network considering all the physical constraints of the distribution network. 
\section{Simulation Results}
\begin{figure}
	\centering
	\includegraphics[width=0.95\columnwidth]{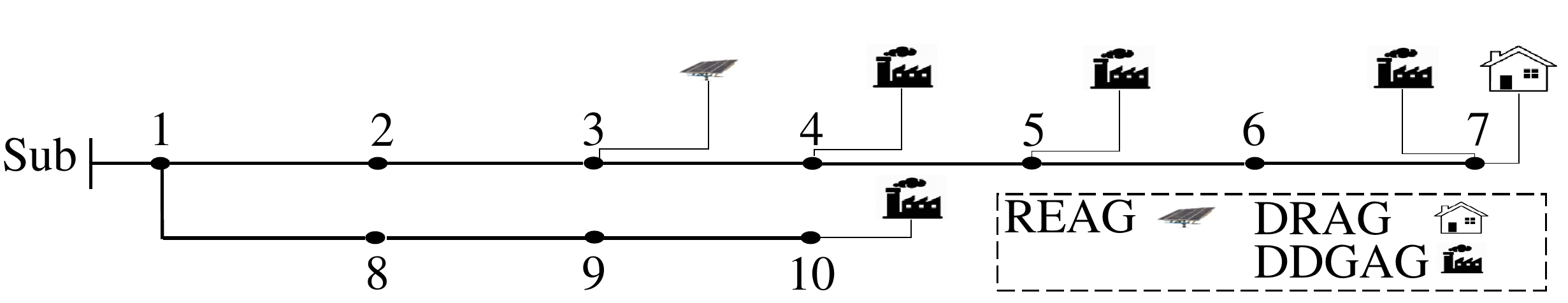}
	\caption{The small distribution network for case studies.}\label{fig.1.distribution network}
\end{figure}
\begin{table}
	\centering
	\caption{Wholesale market participants information}\label{table.1 input data}
	\begin{tabular}{c|c|c|c}
		\hline

		Participant&Pmin (MW)&Pmax (MW) &Offering price (\$/MWh)\\
		\hline
		\hline
		Gen 1&0&10 &8\\
		\hline
		Gen 2&0&20 &20\\
		\hline
		Gen 3&0&30 &22\\
		\hline	
		DR 1&0&10 &30\\
		\hline	
		DR 2&0&20 &32\\
		\hline		
		DR 3&0&20 &34\\
		\hline
	\end{tabular}
\end{table}
\begin{table}
	\centering
	\caption{DSO market participants information}\label{table.2 input data}
	\begin{tabular}{c|c|c|c}
		\hline
		
		Participant&Pmin (MW)&Pmax (MW) &Offering price (\$/MWh)\\
		\hline
		\hline
		DDGAG 1&0&0.5 &20\\
		\hline
		DDGAG 2&0&1 &10\\
		\hline
		DDGAG 3&0&1.2 &15\\
		\hline	
		DDGAG 4&0&2 &24\\
		\hline		
		DRAG &0&20 &28\\
		\hline
	\end{tabular}
\end{table}
The case studies are implemented on a small system containing the ISO running wholesale-level ED and a small distribution network operated by the DSO shown in Fig. \ref{fig.1.distribution network}. In the wholesale-level ED, three generating units, three demand response units, and a firm load of 5 MW is considered. The generating units (Gen) and demand response (DR) units information is in Table. \ref{table.1 input data}. The distribution system in Fig. \ref{fig.1.distribution network} includes 10 nodes, 9 lines, 4 dispatchable distributed generation aggregators (DDGAG), 1 renewable energy aggregators (REAG), and 1 demand response aggregator (DRAG). The distribution system market participants' information is in Table. \ref{table.2 input data}. The REAG production is considered to be 1 MW with no cost. It is assumed that REAG production is constant.

\subsection{The Ideal Case}
\begin{table}
	\centering
	\caption{ISO market outcomes in the ideal case}\label{table.3}
	\begin{tabular}{c|c|c|c}
		\hline
		
		Participant&Share (MW)&Participant&Share (MW)\\
		\hline
		\hline
		Gen 1&10&DDGAG 1 &0.5\\
		\hline
		Gen 2&20&DDGAG 2&1\\
		\hline
		Gen 3&13.8&DDGAG 3&1.2\\
		\hline	
		DR 1&10&DDGAG 4&0\\
		\hline		
		DR 2&20&DRAG 2&2.5\\
		\hline
		DR 3&10&&\\
		\hline
	\end{tabular}
\end{table}
In this section, the simulation results are obtained using the model presented by (\ref{equ.1}) and (\ref{equ.2}). In this case, the DERs participate in the wholesale market directly and submit their offering bids directly to the ISO. This case is ideal since the ISO oversees all the market participants' operation constraints as well as transmission and distribution network constraints. This case is the best case for secure and optimal market participation of the DERs. However, this is not implementable with the current wholesale market structures. 
The market share of each market participant in this model is given in Table. \ref{table.3}. 
\subsection{Participation through the DSO}
\begin{figure}
	\centering
	\includegraphics[width=0.95\columnwidth]{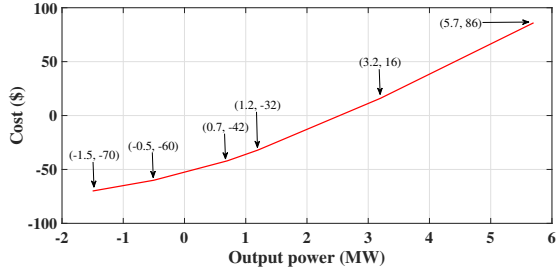}
	\caption{DSO total (minimal) operating cost}\label{fig.2.breakpoints}
\end{figure}

\begin{figure}
	\centering
	\includegraphics[width=0.95\columnwidth]{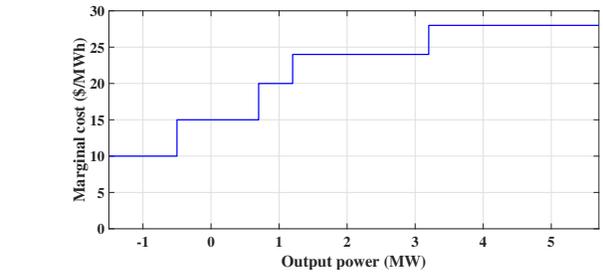}
	\caption{DSO marginal cost function (price-quantity pairs) submitted to ISO}\label{fig.2.DSObid}
\end{figure}

\begin{table}
	\centering
	\caption{ISO market outcomes in the ISO-DSO coordination case}\label{table.4}
	\begin{tabular}{c|c|c|c}
		\hline
		
		Participant&Share (MW) & Participant&Share (MW)\\
		\hline
		\hline
		Gen 1&10 & DR 1&10\\
		\hline
		Gen 2&20 & DR 2&20\\
		\hline
		Gen 3&13.8 & DR 3&10\\
		\hline	
		DSO&1.2& & \\
		\hline
	\end{tabular}
\end{table}
In this case, the bid-in cost function of the DSO is first determined based on the formulation in (\ref{equ.5}). The DSO's total (minimal) operating costs at different output power levels are shown in Fig. \ref{fig.2.breakpoints}. The breakpoints in Fig. \ref{fig.2.breakpoints} are determined by the retail market participants' minimum and maximum output power. The Bid-in marginal cost function (price-quantity pairs offered by the DSO to ISO, which is the derivative of the DSO operating cost curve in Fig. \ref{fig.2.breakpoints}) is in Fig. \ref{fig.2.DSObid}. The bid-in marginal cost function starts with the output power of -1.5 MW which means that DSO can consume energy of 1.5 MW since the DRAG has the (consumption) capability of 2.5 MW and the REAG produces 1 MW. The bid-in price of this consumption is 10 \$/MWh. This is due to the fact that the cheapest unit's price in the DSO market is 10 \$/MWh which means that the wholesale market price should be lower than this value in order for the DSO to buy energy from the wholesale market otherwise it provides that energy from the DDGAG 2. The DSO buys energy with this cost until the capacity of the DDGAG 2 is reached. Then, DDGAG 3, which is the next cheapest generating unit starts to be dispatched. This procedure continues until the last (most expensive) DDGAG is dispatched, which occurs at 3.2 MW. In the end, the DSO sells energy to the wholesale market at the price of 28 \$/MWh which is actually the offering price of the DRAG (for energy consumption). This is due to the fact that if the offering price of the wholesale market is greater than 28 \$/MWh, the DSO sells the energy to the ISO instead of to the DRAG. 
\begin{table}
	\centering
	\caption{DSO market outcomes in the ISO-DSO coordination case}\label{table.5}
	\begin{tabular}{c|c|c|c}
		\hline
		
		Participant&Share (MW) & Participant&Share (MW)\\
		\hline
		\hline
		DDGAG 1&0.5 & DDGAG 3&1.2\\
		\hline
		DDGAG 2&1 & DDGAG 4&0\\
		\hline
		DRAG &2.5 & & \\
		\hline		
	\end{tabular}
\end{table}

The DSO submits this bid-in marginal cost function to the ISO. Then, ISO runs the wholesale market and determines the wholesale market share of the DSO and other participants. The ISO market outcomes are shown in Table. \ref{table.4}. By comparing Tables. \ref{table.3} and \ref{table.4}, it is clear the market outcomes for generating units (Gen) and demand response units (DR) directly participating in the ISO market are identical in the ideal case and the ISO-DSO coordination case.
The share of the DSO is 1.2 MW. In order to determine the share of the market participants in the DSO market, we need to substitute the parameter in the parametric optimization given in (\ref{equ.5}) which results in a simple optimization problem. The results of this optimization problem are given in Table. \ref{table.5}. By comparing Tables. \ref{table.3} and \ref{table.5}, it is obvious the market outcomes for various aggregators are identical in the ideal case (when participating in the ISO market directly) and the ISO-DSO coordination case (when participating in the ISO market through the DSO).
		
\section{Conclusion}
In this paper, an ISO-DSO market coordination framework is proposed to leverage the wholesale market participation of DER aggregators based on parametric programming. The DSO builds the bid-in cost function based on the DSO market participants' offering prices considering their operational constraints and the physical constraints of the distribution network including the power balance equations and voltage limitation constraints. The simulation results performed on the small system indicate the proposed coordination model will result in the same market outcomes as the ideal case in which the DER aggregators directly participate in the wholesale market. 
%
%
%
%

\bibliographystyle{IEEEtran}
\bibliography{Refs}


\end{document}